\documentstyle[prl,preprint,aps,epsfig]{revtex}

\begin{document}
\draft
\tighten

\title{ Structure effects on the Coulomb dissociation of $^8$B  
at relativistic energies}


\author{R. Shyam\P , K. Bennaceur\dag , J. Oko{\l}owicz\dag\ddag ~ and 
M. Ploszajczak\dag}
\address{\P\ Saha Institute of Nuclear Physics, 1/AF Bidhan Nagar, 
Calcutta - 700 064, India}
 \address{\dag\ Grand Acc\'{e}l\'{e}rateur National d'Ions Lourds (GANIL),
CEA/DSM -- CNRS/IN2P3, BP 5027, F-14076 Caen Cedex 05, France}
\address{\ddag\ Institute of Nuclear Physics, Radzikowskiego 152,
PL - 31342 Krak\'{o}w, Poland}

\maketitle  
\begin{abstract}
We investigate the Coulomb dissociation of $^8$B on $^{208}$Pb target  at
the beam energy of 250 MeV/nucleon, employing the cross sections
for the radiative capture reaction $^7$Be(p,$\gamma$)$^8$B calculated
within the Shell Model Embedded in the Continuum (SMEC) approach.
In contrast to the situation at lower beam energies, the Coulomb breakup
cross sections are found to be sensitive to the $M1$ transitions. 
Comparisons of SMEC and single-particle potential model predictions show that
the Coulomb breakup cross sections at these high energies are 
sensitive to the structure model of $^8$B. Comparison  
with the preliminary data on the angle integrated spectra
reported recently by the GSI group shows that the theory is able to 
reproduce the absolute magnitude as well as the shape of this
data well. The contributions of the $E2$ component strongly depends 
on the range of the angles of the $\mbox{p}-{^7}\mbox{Be}$ center of
mass with respect to the beam direction, included 
in the angular integrations of the double differential cross section.
If integrations are done over
the angular range of 0$^\circ$-1.87$^\circ$, the $E2$ multipolarity
contributes up to 25 $\%$ to the cross sections even for
the relative energies of [$\mbox{p} - {^7}\mbox{Be}$] below 0.25 MeV.
However, these contributions are reduced by an order of 
magnitude at lower relative energies if the maximum of the angle integration
is $\sim$ 1$^\circ$.     
\end{abstract}

\bigskip
\noindent

\vfill
\newpage

\section{Introduction}       
The $^8$B isotope produced in the Sun via the radiative capture reaction
$^7$Be(p,$\gamma$)$^8$B
is the principal source of the 
high energy neutrinos detected in the Super-Kamiokande and $^{37}$Cl
detectors~\cite{bahc89}. In fact the calculated rate of events in the
former detector (and also in the SNO experiment~\cite{bahc98})  
is directly proportional to the rate of this reaction measured 
in the laboratory at low energies ($\sim$ 20 keV)
\cite{bahc98}. Unfortunately, the measured
cross sections (at relative energies ($E_{CM}$) of 
[$\mbox{p} - {^7}\mbox{Be}$] $>$ 200 keV)
disagree in absolute magnitude and the value extracted by extrapolating
the data in the region of 20 keV differ from each other by 30-40 $\%$.
This makes the rate of the reaction $^7$Be($p,\gamma)^8$B most poorly
known quantity in the entire nucleosynthesis chain leading to
the formation of $^8$B~\cite{adel98}.

The Coulomb dissociation (CD) method provides an alternative indirect
way to determine the cross sections for the radiative capture reactions
at low energies~\cite{baur86,baur94,shya96,shya97,baur97,shya99}.
In this method, the 
radiative capture is reversed by the dissociation of the projectile
(the fused system) in the Coulomb field of the target by assuming that the
strong interaction between the nuclei is absent and the electromagnetic
excitation process is dominated by a single multipolarity
(see {\it e.g.} Refs. \cite{baur94,baur97}).
However, in the CD of $^8$B, the contributions of
$E2$ and $M1$ multipolarities as well as nuclear breakup can be
disproportionately enhanced in certain kinematical
regimes~\cite{lang94,gai95} and a careful
investigation~\cite{shya99,gai98} is necessary to isolate the
conditions in which these terms have negligible effect on the
calculated breakup cross sections.

For the CD measurements of $^8$B performed at
RIKEN~\cite{moto94,kiku97,kiku98} at around 50 MeV/nucleon, it was
found in a theoretical model\cite{shya99} that the data are free from
nuclear and $E2$ contributions if measurements are limited to
$E_{CM}$ $<$ 0.70 MeV and to angles of $p-^7$Be c.m. with respect
to the beam direction ($\theta_{CM}$) $<$4$^\circ$. On the other hand, for
measurements at subCoulomb
energies~\cite{joha96}, these contributions turned out to be dominant
everywhere~\cite{shya99,nune99}.
Furthermore, the multi-step breakup effects were also found
to be quite important at these energies \cite{nune99,type97}

Recently, the Coulomb dissociation of $^8$B on a $^{208}$Pb target
has been performed at GSI-Darmstadt at the beam energy
of 250 MeV/nucleon~\cite{boue99,iwasa99}. The advantages of the
CD process at this high energy are:  
(i) measurements at $E_{CM}$ lower than RIKEN are possible   
due to certain experimental advantages~\cite{boue99,iwasa99}.
(ii) the multi-step breakup processes  
(${\it e.g.}$ the Coulomb post-acceleration) are
negligible~\cite{type97,bert95}, and
(iii) the nuclear breakup processes are expected to be negligible in 
comparison to the Coulomb one due to the strong enhancement in the virtual
photon spectrum~\cite{baur97,bert95}, an advantage that existed already
in the experiments performed at the RIKEN energies in the kinematical
regime as mentioned above~\cite{shya99}. At the beam energy of
250 MeV/nucleon, the $E2$ component is expected to be about one
third of that found at RIKEN energies although, at the same time,
the $E1$ component is also reduced by about one half. But more
importantly, the $M1$ component is expected to be enhanced particularly
in the first resonance region (0.5 MeV $<$ $E_{CM}$ $<$ 0.65 MeV).
This provides an alternative opportunity to test various models of
the $^8$B structure, as they differ in their treatment of the resonance
structure of this nucleus~\cite{csoto99,nune97,fedo95,desco94}. 

The aim of this paper is to perform calculations of the CD of
$^8$B on $^{208}$Pb target at the beam energy of 250 MeV/nucleon,
using the cross section for the cature reaction
$^7$Be(p,$\gamma$)$^8$B calculated within 
a recently developed realistic shell model
which includes coupling between many
particle quasi-bound states and the continuum of one particle
scattering states (to be referred as the Shell Model Embedded
into the Continuum (SMEC) approach) \cite{bnop1,bnop2}.
In order to investigate the sensitivity of the CD process to the 
structure of $^8$B, we also perform calculations within 
a single particle potential model
using the parameters ginen by Esbensen and Bertsch (EB)~\cite{esbe96}.
We compare the results of our calculations with the preminary data
on the angle integrated cross sections for this reaction 
taken recently in an experiment performed at the GSI,
Darmstadt~\cite{boue99,iwasa99}. 
We discuss the kinematical regime where the breakup measurements
performed at these energies can be reliably
used to extract the astrophysical $S$-factor ($S_{17}$)
for the $^7$Be($p,\gamma)^8$B capture reaction.

The remainder of this paper is organized in the following way.
The formalism of the SMEC approach is described in the next section. 
The formulas used in the calculation of the Coulomb dissociation 
cross sections are also described here. The results of our numerical 
calculations and their discussions are presented in Sect. 3, while the
summary and conclusions of our work is given in Sect. 4.

\section{The formalism}

\subsection{The shell model embedded in the continuum (SMEC)}
In the next chapter, we shall study the CD process  
$^{8}\mbox{B} \rightarrow {^{7}\mbox{Be}} + p$, using the Shell Model Embedded 
in the Continuum (SMEC) which has been applied recently to examine
structure  for mirror nuclei, $^{8}\mbox{B}$,
$^{8}\mbox{Li}$ , and capture cross sections for mirror reactions   
$^7\mbox{Be}(p,\gamma)^8\mbox{B}$, 
$^7\mbox{Li}(n,\gamma)^8\mbox{Li}$ \cite{bnop1,bnop2}. The SMEC model, in which
realistic Shell Model (SM) solutions for (quasi-)bound states are coupled to the
one-particle scattering continuum,  is a development of the Continuum
Shell Model (CSM) \cite{bartz1,bartz2,bartz3} 
for the description of low excitation energy
properties of weakly bound nuclei. In the SMEC, the bound
(interior) states together with its environment of asymptotic scattering 
channels form a quantum closed system. Using 
the projection operator technique, one separates the $P$ subspace of
asymptotic channels from the $Q$ subspace of many-body states which are
build up by the bound s.p. wave functions
and by the s.p.\ resonance wave functions. $P$ subspace  is assumed to
contain $(N-1)$ - particle states with nucleons on bound s.p.\ orbits
and one nucleon in the scattering state. Also the s.p.\
resonance wave functions outside of the cutoff radius $R_{cut}$~ are
included in the $P$ subspace. The resonance wave functions for 
$r < R_{cut}$~, are included in the $Q$ subspace. The
wave functions in $Q$ and $P$ are then properly renormalized
in order to ensure the orthogonality of wave functions in both subspaces. 

For the (quasi-) bound many-body states in $Q$ subspace 
one solves the SM problem 
\begin{eqnarray}
\label{eq1}
H_{QQ}{\Phi}_i = E_i{\Phi}_i
\end{eqnarray}
where $H_{QQ} \equiv QHQ$~ is 
the SM effective Hamiltonian. 
What should be taken for coupling between bound and scattering
states is in principle not known and we have decided to use a schematic
combination of Wigner and Bartlett forces \cite{bnop2} 
\begin{eqnarray}
\label{force}
V_{12} = -V_{12}^{(0)}
[\alpha + (1-\alpha )P_{12}^{\sigma}]\delta({\bf r}_1 - {\bf r}_2)
\end{eqnarray}
where $P_{12}^{\sigma}$ is the spin exchange term and $(1 - \alpha )$ is the
spin exchange parameter.
We assume that effects of this coupling on the effective interaction in $Q$
subspace are already contained in $H_{QQ}$.

The SM wave function has an
incorrect asymptotic radial behavior for unbound states.
Therefore, to generate both the radial s.p.\ wave functions in the $Q$ subspace
and the scattering wave functions in $P$ subspace
we use the average potential of Saxon-Woods (SW) type
with the spin-orbit part included
\begin{eqnarray}
\label{pot}
U(r) = V_0f(r) + V_{SO} (4{\bf l}\cdot{\bf s})
\frac{1}{r}\frac{d{\rho}(r)}{dr}  + V_C  
\end{eqnarray}
where $f(r)$ is the spherical symmetric SW form factor
\begin{eqnarray}
\label{pot1}
{\rho}(r) = \left[ 1 + \exp \left( \frac{r-R_0}{a} \right) \right]^{-1} ~ \ .
\end{eqnarray}
The Coulomb potential $V_C$ is calculated for the uniformly charged sphere
with radius $R_0$.

For the continuum part, one solves the coupled channel equations
\begin{eqnarray}
\label{esp}
(E^{(+)} - H_{PP}){\xi}_{E}^{c(+)} \equiv 
\sum_{c^{'}}^{}(E^{(+)} - H_{cc^{'}}) {\xi}_E^{c^{'}(+)} = 0 
\end{eqnarray}
where index $c$~ denotes different channels and $H_{PP} \equiv PHP$~. 
The superscript $(+)$ means that boundary
conditions  for outgoing scattering waves are used.
The channel states are defined by coupling one
nucleon in the scattering continuum to a many-body state of $(N - 1)$ - nucleus 
given by the SM. The channel - channel coupling potential in (\ref{esp}) is 
\begin{eqnarray}
\label{esp2}
H_{cc^{'}} = (T + U ){\delta}_{cc^{'}} + {\upsilon }_{cc^{'}}^{J} 
\end{eqnarray}
where $T$ is the kinetic-energy operator and
$ {\upsilon }_{cc^{'}}^{J}$ is the channel-channel coupling generated by the
residual interaction. $U$
in (\ref{esp2}) consists of an 'initial guess' $U(r)$ given by (\ref{pot}) 
and the diagonal part of the coupling potential ${\upsilon }_{cc}^{J}$
which depends on both the s.p.\ orbit
${\phi}_{l,j}$ and the considered many-body
state $J^{\pi}$. Obviously, this
correction cannot be neglected when generating s.p.\ wave function
${\phi}_{l,j}$ for a given $J^{\pi}$. These s.p.\ wave functions
define $Q$ subspace and thus modify the diagonal part of the residual force.
This implies  a self-consistent iterative
procedure, because the change of s.p.\ wave
function changes the correction coming from the residual force, and so on. This
procedure yields the new {\it self-consistent potential} 
\begin{eqnarray}
\label{usc}
U^{(sc)}(r) = U(r)+{\upsilon }_{cc}^{J(sc)}(r)
\end{eqnarray}
and consistent with it the
new {\it renormalized matrix elements} of the
coupling force. The parameters of the initial potential $U(r)$
are chosen  in such a way that $U^{(sc)}(r)$
reproduces energies of experimental s.p.\ states,
whenever their identification is possible. 

The third system of equations in SMEC consists of inhomogeneous
coupled channel equations
\begin{eqnarray}
\label{coup}
(E^{(+)} - H_{PP}){\omega}_{i}^{(+)} = H_{PQ}{\Phi}_i \equiv w_i
\end{eqnarray}
with the source term $w_i$ which is primarily given by the 
structure of $N$ - particle SM wave function ${\Phi}_i$~. The explicit
form of this source was given in \cite{bnop2}.
These equations define functions ${\omega}_{i}^{(+)}$~,  which
describe the decay of quasi-bound state ${\Phi}_i$~ in the continuum.
The source $w_i$~ couples the wave function of $N$ - nucleon
localized states with $(N-1)$ - nucleon localized states plus one nucleon
in the continuum. Form factor of the source term is given by 
the self-consistently determined s.p.\ wave functions.

The full solution is expressed by three functions
${\Phi}_i$~, ${\xi}_{E}^{c}$~ and ${\omega}_i$ \cite{bnop2,bartz1}~
\begin{eqnarray}
\label{eq2}
{\Psi}_{E}^{c} & = & {\xi}_{E}^{c} + \nonumber \\ & + & 
\sum_{i,j}({\Phi}_i + {\omega}_i)
\frac{1}{E - H_{QQ}^{eff}}
<{\Phi}_{j}\mid H_{QP} \mid{\xi}_{E}^{c}> 
\end{eqnarray}
where
\begin{eqnarray}
\label{eq2a}
H_{QQ}^{eff} = H_{QQ} + H_{QP}G_{P}^{(+)}H_{PQ} 
\end{eqnarray}
is the {\it effective}
SM Hamiltonian which includes the coupling to the continuum, and
$G_{P}^{(+)}$~ is the Green function for the motion of s.p.\ in
the $P$ subspace. Matrix $H_{QQ}^{eff}$ is non-Hermitian
(the complex, symmetric matrix)  for energies above the particle
emission  threshold and Hermitian (real) for lower energies. 
The eigenvalues, ${\tilde {E_i}} - \frac{1}{2}i{\tilde {{\Gamma}_i} }$ , are
complex for decaying states and
depend on the energy $E$ of particle in the continuum.
The energy and width of resonance states is determined by the condition 
$\tilde{E_i}(E) = E$ \cite{bartz1}. Inserting them in (\ref{eq2}), one 
obtains 
\begin{eqnarray}
\label{cons}
{\Psi}_{E}^{c} & = & {\xi}_{E}^{c} + \nonumber \\ & + &  
\sum_{i}^{}{\tilde {\Omega}_i}
\frac{1}{E - {\tilde E_i}
+ (i/2){\tilde {\Gamma}_i}} <{\tilde {\Phi}_i} \mid H \mid {\xi}_{E}^{c}>
\end{eqnarray}
for the continuum many-body wave function projected on channel $c$ , where
\begin{eqnarray}
\label{diss}
{\tilde {\Omega}_i} & = & {\tilde {\Phi}_i} + \nonumber \\ & + & \sum_{c}
\int_{{\varepsilon}_c}^{\infty} dE^{'} {\xi}_{E^{'}}^{c}
\frac{1}{E^{(+)} - E^{'}}
<{\xi}_{E^{'}}^{c}\mid H \mid {\tilde {\Phi}_i}> 
\end{eqnarray}
is  the wave function of discrete state modified by the coupling to the
continuum states. It should be stressed that the 
SMEC formalism is {\it fully symmetric} in treating the
continuum and bound state parts of the solution, ${\Psi}_{E}^{c}$ 
represents the continuum state modified by the discrete states, and
${\tilde {\Omega}_i}$ represents the discrete
state modified by the coupling to the continuum. 

\subsection{SMEC wave functions for $^{8}\mbox{B}$}

 The SMEC results depend mainly on 
(i) the effective nucleon - nucleon interaction in $Q$ subspace,
(ii) the residual coupling of
$Q$ and $P$ subspaces, (iii) the self-consistent
average s.p.\ potential which generates the radial
form factor for s.p.\ bound wave functions and s.p.\ resonances.
Cohen - Kurath (CK) interaction \cite{cohen} is used for the SM effective 
interaction in $Q$ subspace. The freedom of choosing
asymptotic conditions in solving eqs. (\ref{coup}) means
that the zero on the excitation energy scale can be fixed
arbitrarily. We choose the zero on this scale by requiring  
that the lowest resonance ($J^{\pi}=1_{1}^{+}$)
with respect to the proton emission threshold has its
experimental position.         
An essential element of SMEC approach is the construction of $Q$ -
subspace. This is achieved by an iterative procedure which
yields the self-consistent s.p.\ potential depending on the
s.p.\ wave function ${\phi}_{l,j}$, the total spin $J$ of the $N$-nucleon
system as well as on the one-body matrix elements of $(N-1)$ - nucleon
daughter system. The parameters of initial SW potentials for different
contributions of spin exchange component in the residual interaction
(\ref{force}) are summarized in Table I. 

The unique
potential $U(r)$ is used for the calculation of self-consistent
potentials for {\it all many-body states} in $^{8}\mbox{B}$, and for both
$1p_{3/2}$ and $1p_{1/2}$ proton s.p.\ states.
For neutrons, there is no correction from the residual interaction,
and the average s.p.\ potential is chosen such that it
yields $1p_{3/2}$  and $1p_{1/2}$ neutron orbits
at $-13.02\,$MeV and  $-11.16\,$MeV respectively \cite{bnop2}. 

The quadrupole moment $<Q>$ of $^{8}\mbox{B}$ is a useful test of
the SMEC wave function. We have calculated
$<Q>$ following the approach of Carchidi et al. \cite{carchidi}.  
For the residual force (\ref{force}) with 
$V_{12}^{(0)}$ = 650 MeV$\cdot$fm$^{3}$ and for the spin exchange parameter
$(1 - \alpha ) = 0.05$ one finds \cite{bnop2}, $<Q>=6.99 ~\mbox{e fm}^2$, 
in the good agreement with the
experimental value \cite{minamisono}, $<Q> = 6.83 \pm 0.21 ~\mbox{e fm}^2$.
This theoretical value has been obtained assuming the effective charges,  
$e_p=1.35$ , $e_n=0.35$, and the SM spectroscopic 
factors for the CK interaction.

SMEC results depend sensitively on very
small number of parameters. Some of them, like the parameterization of the
residual interaction which couples states in $Q$ and $P$ subspaces, has been
established previously \cite{bnop1,bnop2}. 
The others, related to the energy of
s.p.\ states which determine the radial wave function of many-body states, are
bound by the SM spectroscopic factors and experimental binding energy in
studied nuclei. The spectrum of $^{8}\mbox{B}$ depends strongly on
couplings to the ground state (g.s.) 
of $^{7}\mbox{Be}$ but changes very little if also the
couplings to the excited state $3/2^{-}$ of
$^{7}\mbox{Be}$ are taken into account \cite{bnop2}. 
Width for $J^{\pi}=1_{1}^{+}$ state depends
sensitively on the proportion of direct and
spin exchange terms (see Table II) in the residual
interaction. 

Varying the
parameter of the spin exchange component for a fixed
intensity of the coupling, we came to the conclusion that most satisfactory
description of experimental data are achieved for small contribution of the
spin exchange part, {\it i.e.}, approaching the limit of pure Wigner force
\cite{wigner}. This finding is consistent with the results of
SM which strongly suggest an approximate validity of $SU(4)$
symmetry in $p$-shell nuclei \cite{cohen,nang,mukhopadhyay,john}.

SM energy of the first $J^{\pi}=3_{1}^{+}$ level is
too low as compared to the experimental value (see 
Table II).                          
The coupling to the continuum cannot correct for this deficiency.
The width of $3_{1}^{+}$ state differs by at least a factor 5
from the experimental data and here, again, the agreement between experiment
and calculations improves when $(1 - \alpha ) \rightarrow 0$ .
There are several reasons for this 
discrepancy. Firstly, SM with CK interaction is not well describing energy of
this state and, as pointed above, the width of the many-body state depends on
its excitation energy with respect to the particle emission threshold (see 
Table II).  Secondly, the wave function of
experimental $3_{1}^{+}$ state is certainly overlapping with the cluster
configuration [${^{3}\mbox{He}}-{^{5}\mbox{Li}}$], which cannot be adequately
described in $p$-space SM calculations. Experimental 
$3_{1}^{+}$ state lies above the threshold for three-particle decay,
$^{8}\mbox{B} \longrightarrow [{^{3}\mbox{He}}-\mbox{p}-{^{4}\mbox{He}}]$.
This decay channel largely contributes to the $3_{1}^{+}$ width, but
cannot be accounted for in the approximation of one-particle 
scattering continuum. We have to keep in mind these limitations
when analyzing the Coulomb dissociation cross-section for $^{8}\mbox{B}$ .

\subsection{Radiative capture }
The calculation of the capture cross-section in the SMEC 
goes as follows. The initial wave function for 
[$\mbox{p} \bigotimes {}^{7}\mbox{Be}$] system is 
\begin{eqnarray}
\label{psiin}
\Psi_i(r)=\sum_{l_a j_a}i^{l_a}{\psi_{l_a j_a}^{J_i}(r)\over r}
\biggl[\bigl[Y^{l_a}\times\chi^{s}\bigr]^{j_a}\times\chi^{I_t}\biggr]^{(J_i)}
_{m_i} 
\end{eqnarray}
and the final wave function for 
$^{8}\mbox{B}$ in the g.s. ($J^{\pi}=2^{+}$) is 
\begin{eqnarray}
\label{psifin}
\Psi_f(r)=\sum_{l_b j_b}A_{l_bsj_b}^{j_bI_bJ_f} {u_{l_b j_b}^{J_f}(r)\over r}
\biggl[\bigl[Y^{l_b}\times\chi^{s}\bigr]^{j_b}\times\chi^{I_t}\biggr]^{(J_f)}
_{m_f}  
\end{eqnarray}
$I_t$~ and $s$~ denote the spin of target nucleus and
incoming proton, respectively.
$A_{l_bsj_b}^{j_bI_bJ_f}$~ is the coefficient of fractional parentage and
$u_{l_bj_b}^{J_f}$~ is the s.p.\ wave in the many-particle state~$J_f$~.
These SMEC wave functions, ${\Psi}_i(r)$~ , ${\Psi}_f(r)$~, are then used to
calculate the transition amplitudes $T^{E{\cal L}}$ and $T^{M1}$ for
$E1$~, $E2$~ and $M1$ transitions, respectively \cite{bnop1,bnop2}.
The radiative capture cross section is 
\begin{eqnarray}
\label{tran}
\sigma^{E1,M1} & = & {16\pi\over9} \biggl({k_\gamma\over k_p}\biggr)^3
  \biggl({\mu\over\hbar c}\biggr)
  \biggl({e^{2}\over\hbar c}\biggr)
  {1\over 2s+1}~{1\over 2I_t+1} \times \nonumber \\ & \times &
\sum\mid T^{E1,M1}\mid^2
\end{eqnarray}
\begin{eqnarray}
\label{tran1}
\sigma^{E2} & = & {4\pi\over75} \biggl({k_\gamma^5\over k_p^3}\biggr)
  \biggl({\mu\over\hbar c}\biggr)
  \biggl({e^{2}\over\hbar c}\biggr)
  {1\over 2s+1}~{1\over 2I_t+1} \times \nonumber \\ & \times &
\sum\mid T^{E2}\mid^2  
\end{eqnarray}
where $\mu$~ stands for the reduced mass of the system.

The astrophysical S-factor ($S_{17}$) is related to the capture cross
section by
\begin{eqnarray}
S_{17}(E_{CM}) & = & \sigma^{\pi L}(E_{CM})\,E_{CM}\,e^{[2\pi \eta(E_{CM})]},
\end{eqnarray}
where $\eta(E_{CM})$ is the Coulomb parameter and $\pi L$ represents the 
multipolarity.
\subsubsection{SMEC results for $^{7}\mbox{Be}(p,\gamma )^{8}\mbox{B}$}

In Fig. 1, we show the astrophysical S-factors ($S_{17}$) (as defined 
by Eq. (2.17)) of $E1$ (solid lines),
$E2$ (dashed-dotted line) and $M1$ (dashed line) 
multipolarities as a function of
c.m.\ energy for four versions of SMEC used in the calculations 
performed in this paper (see also Fig. 7 of Ref. \cite{bnop2}). The parts
(a)-(d) show the results obtained with versions I,II,III,IV of the
SMEC model respectively. 
Version I corresponds to the spin exchange parameter $(1-\alpha)$
in the residual coupling (\ref{force}) equal $0.27$. This is the
standard value resulting from a fit to the giant dipole resonance
in $^{16}\mbox{O}$ \cite{bartz1,buck}. In version II,
$(1-\alpha )=0.05$, that corresponds to an
almost pure Wigner force limit for this coupling. Strength of the residual
interaction in versions I and II of the SMEC model is,  
$V_{12}^{(0)}$ = 650 MeV$\cdot$fm$^{3}$ and
the parameters of the initial SW potentials are given in Table I.
Versions III and IV are the same as versions I and II respectively except that
the resonant $M1$ contribution for the $J^{\pi}=3_{1}^{+}$ state has been
omitted. This resonance has much smaller width than seen in the data due
mainly to the missing three-body 
final state in the continuum, as  explained in Sect. II.B.              
The $M1$ contribution and particularly its resonant part, 
are strongly dependent on the spin exchange parameter 
$(1-\alpha )$ in the residual coupling (\ref{force}). For $(1 - \alpha ) =
0.05$ the resonant part of $M1$~ transitions
yields $S^{M1}=20.52$ eV${\cdot}$b at the $1_{1}^{+}$
resonance energy. This value is proportional to the square of spectroscopic 
amplitude of $p$-states, which for the CK interaction is
$-0.352$ and 0.567 for $p_{1/2}$ and $p_{3/2}$ respectively. 
Ratio of $E2$ and $E1$ contributions at the position of $1_{1}^{+}$
resonance is $8.15\cdot$$10^{-4}$ or $7.72\cdot$$10^{-4}$
depending on whether $(1-\alpha)$ equals 0.27 or 0.05.
The experimentally deduced value for this ratio,
$6.7_{-1.9}^{+2.8}\cdot 10^{-4}$ \cite{davids}, is consistent with both values
of $(1-\alpha )$. The $E2$ contribution contains both resonant and non-resonant
contributions. Their ratio at $1_{1}^{+}$ resonance is 0.187 
(for $(1 - \alpha ) = 0.05$), and the contribution from different
initial states, $0^{+}, 1^{+}, 2^{+}, 3^{+}$ and $4^{+}$ is
1.17$\cdot$10$^{-5}$ $\mu$b, 
7.2$\cdot$10$^{-5}$ $\mu$b, 4.64$\cdot$10$^{-5}$ $\mu$b, 
6.01$\cdot$10$^{-5}$ $\mu$b  and 5.64$\cdot$10$^{-5}$ $\mu$b, respectively.
Importance of this quantity has been suggested recently by Barker
\cite{barker1}.

$E1$ component provides the main 
contribution to the total capture cross-section.
The low energy behavior of the astrophysical factor $S_{17}(E)$ can be
approximated by,
$S_{17}(E) = S_{17}(0) \exp ({\hat \alpha}E+{\hat \beta}E^{2})$.
In the range of c.m.\ energies up to 100 keV, it yields 
$S_{17}(0)=19.594$ eV$\cdot$b,  ${\hat \alpha}=-1.544\,$MeV$^{-1}$,
${\hat \beta}=6.468\,$MeV$^{-2}$ for $(1-\alpha ) = 0.05$. 
The above value of $S_{17}(0)$ 
is close to the values reported by Filippone et al. \cite{filippone} and
Hammache et al. \cite{hammache}. At the
position of $1_{1}^{+}$ resonance, the calculated
$S_{17}$ - factor ($S_{17} = 40.67$ eV${\cdot}$b)
is somewhat smaller than that reported by Filippone et al. \cite{filippone} .

\subsection{Single-particle model for $^{8}\mbox{B}$}

For a comparison, we shall also study the CD of  $^{8}\mbox{B}$
using a simple s.p. description of loosely bound proton in
$^{8}\mbox{B}$, using the potential parameters given by Esbensen
and Bertsch (EB)\cite{esbe96}. This will be referred to as the EB 
potential model in the following. The S-factors obtained with this model
are shown by lines with solid circles in part (c) of Fig. 1.
In this model, both the
g.s. $2_{1}^{+}$ as well as the resonances $1_{1}^{+}$ and  $3_{1}^{+}$  are
assumed to have the structure $[^{7}\mbox{Be}(3/2^{-}) \otimes 1p_{3/2}]$, {\it
i.e.}, the spectroscopic amplitudes for these states are assumed to be equal 1.
This assumption is perhaps questionable for the $1_{1}^{+}$ and  $3_{1}^{+}$ 
resonances \cite{barker} (the experimental spectroscopic factor
for $1_{1}^{+}$ resonance for the mirror state in $^{8}\mbox{Li}$ is
0.48 \cite{ajzenberg} ). No intrinsic excitations of the $^{7}\mbox{Be}$
are allowed and the s.p. potential well is adjusted for each state
separately to reproduce either the one-proton separation energy (for
the g.s. $2_{1}^{+}$) or the excitation energies (for the $1_{1}^{+}$
and  $3_{1}^{+}$ resonances). This extreme s.p. description of the
$^{8}\mbox{B}$ yields the width of resonances appreciably
bigger than the experimental value, in contrast to the SMEC, which
yields too narrow width for those states. For example, the $1_{1}^{+}$
width in the EB potential model is 70 keV as compared to the measured
value $37 \pm 5$ keV and the $3_{1}^{+}$  width is 1740 keV instead
of 350$\pm$40 keV found experimentally \cite{ajzenberg}. In SMEC widths
of $1_{1}^{+}$ and $3_{1}^{+}$ states are 26 keV and 35 keV, and
16.5 keV and 13 keV for for $(1 - \alpha ) = $ 0.05  and 0.27
respectively~\cite{bnop2}.

In the EB potential model, the ratio  $S^{E2}/S^{E1}$ at the
$1_{1}^{+}$ peak energy equals $9.5\cdot 10^{-4}$ and situates
at the upper most limit of the experimental values deduced by Davids
et al. \cite{davids}. Ratio of resonant and nonresonant $E2$
contributions at this energy is 0.305. The astrophysical factor
$S_{17}(0)$ is $\simeq$ 18.4 eV$\cdot$b and is close to the value in
the SMEC approach.

\subsection{Coulomb dissociation cross section}

The double differential cross-section for the Coulomb excitation
of $^8$B from its g.s. to the continuum, with a
definite multipolarity of order $\pi\lambda$ is given
by~\cite{baur86,baur94,shya96}
\begin{eqnarray}
\label{cdiss}
\frac{d^2 \sigma}{d \Omega_{8_{B^*}} dE_{CM}} & = &
            \sum_{\pi\lambda}\frac{1}{E_{CM} }
                       \frac{dn_{\pi \lambda}}{d \Omega_{8_{B^*}}}
                             \sigma_{\gamma}^{\pi
                              \lambda}(E_\gamma), 
\end{eqnarray}
In Eq. (17) 
$\Omega_{8_{B^*}}$ defines the direction of the c.m. of the $p-^7$Be
system (to be referred as $^8$B$^*$) with respect to the beam direction. 
$\sigma_{\gamma}^{\pi \lambda}(E_{\gamma})$ is the cross-section for
the photo-disintegration process
$\gamma + ^8$B $\rightarrow ^7$Be + $p$,
with photon energy $E_\gamma$, and multipolarity
$\pi\,=\,$ E (electric) or M (magnetic), and
$\lambda\,=\, 1,2...$ (order), which is related to that of
the radiative capture process $^7$Be + $p$ $\rightarrow$
$^8$B + $\gamma$ through the theorem of detailed balance.
$E_\gamma$ is given by $E_{CM} = E_{\gamma} + Q $, with
$Q = 0.137$.     
In most cases, only one or two multipolarities dominate the
radiative capture as well as the Coulomb dissociation cross sections.
$n_{\pi\lambda}(E_\gamma)$ in Eq. (\ref{cdiss}) represents the number of 
equivalent (virtual) photons provided by the
Coulomb field of the target to the projectile, which is
calculated by the methods discussed in Ref.~\cite{shya96,alex89}.

\section{Results and discussion}

In Fig. 2, we show the results of the calculations for the energy
differential cross sections for the reaction
$^8$B + $^{208}$Pb $\rightarrow$ $^8$B$^*$ + $^{208}$Pb
at the beam energy of 250 MeV/nucleon, using the capture cross sections
obtained with versions I, II, III, IV of SMEC. 

Results shown in Fig. 2 have been obtained by integrating
Eq. (\ref{cdiss})) for $\theta_{8_{B^*}}$ angles from
0.01$^\circ$ - 1.87$^\circ$, which is the range of the angle integration 
in the 
preliminary GSI data as reported in~\cite{boue99,iwasa99}.
The cross sections for $E1$, $E2$ and $M1$ multipolarities are shown
by dashed, dotted and dashed-dotted lines respectively, while the solid
line shows their sum. We note that all the four models reproduce
the preliminary GSI data (which has been taken from the thesis of
Federick Boue~\cite{boue99}) well. However, in models I and II, a peak
appears at $E_{CM} \sim 1.2$ MeV. This
is because these models include also the contributions from the
3$^+$ state in the $^8$B continuum which in SMEC occurs at about 1.2
MeV excitation energy. This gives rise to a strong
resonance in the $M1$ capture cross sections which is reflected in the
corresponding Coulomb dissociation results.
In models III and IV, contributions
from this state is not included. In each case, the $E2$ multipolarity
contributes to the extent of about 25\%. The maximum in the 
experimental cross sections is around $E_{CM} = 0.5$ MeV. 
The calculations cannot reproduce this without strong
contributions from the $M1$ multipolarity, as will be shown 
later on.

In Fig. 3, we present the comparison of the Coulomb dissociation 
cross sections (for the same reaction as in Fig. 2), obtained with
capture cross sections of versions III (a) and IV (b) of the SMEC model
and the EB potential model. The  results for $E1$, $E2$
and $M1$ multipolarities for SMEC models and the EB potential model
are shown by dashed-dotted and dotted lines respectively.
Their sum is shown by the solid and dashed lines in the two cases.
We note that the SMEC cross sections are in somewhat better agreement with
the preliminary GSI data. As compared to SMEC, the EB calculations
under-predict the experimental data for $E_{CM}$ larger than
about 0.8 MeV, while they over-predict it around the 1$^+$ 
resonance region. This can be traced back to the differences in
predictions of the two models for the direct capture cross sections of 
various multipolarities (see Fig. 1). 
We note that while the $E1$ cross sections obtained with SMEC 
(version III) and EB model are similar, the two differ in case of SMEC
version IV. On the other hand, SMEC E2 cross sections are always larger
than that of the EB potential model. At the same time, the $M1$ cross
section of the latter are much larger and wider in width
as compared to that of former. 

Therefore, the Coulomb dissociation data
at these high energies seem to show sensitivity to the capture cross 
sections calculated within different models of $^8$B structure. Although
these are only preliminary data, yet SMEC models appear to be in
a somewhat better agreement with it. Unfortunately, it is difficult 
to commment on the the difference seen in the widths of the $M1$
resonance in the two models. This is because the  
experiments have a finite resolution ($\sim$ 110 keV at
$E_{CM}$ = 0.63 MeV~\cite{iwasa99}) that may give a large aparent
width to the cross sections near the resonance region. Furthermore, 
the data are presented in the larger energy bins. It would, 
be worthwhile to improve upon both these aspects in the future
studies. However, it must be added here that the width of this
resonance is tested basically with the direct radiative capture 
$(p,\gamma)$data.  
 
We next investigate the role of the $M1$ multipolarity in the
high energy data of GSI. We would like to recall that at lower
beam energies ({\it e.g.}, the RIKEN experiments~\cite{moto94,kiku97}),
the contribution of this multipolarity was almost negligible. 
In Fig. 4, we show the CD predictions (obtained 
with SMEC, version III) for  $E1$, $E2$ and $M1$ 
components of the angular distributions for the      
$^8$B + $^{208}$Pb $\rightarrow$ $^8$B$^*$ + $^{208}$Pb reaction
at the beam energies of 51.9 MeV/nucleon. For completeness sake we also
show the experimental data of Kikuchi et al.~\cite{kiku97}. Note
that since these CD calculations have been done within a 
pure semi-classical theory~\cite{ald75}, the agreement with the 
data, beyond 4$^\circ$ is not good as compared to that seen in
Ref.~\cite{shya99}. As has been discussed in
Ref.~\cite{shya99}, the point like projectile approximation of the
semi-classical theory breaks down at angles beyond this. Inclusion of
finite-size effects of the projectile reduces the cross sections at
larger angles which leads to a better agreement with the experimental
data \cite{shya99}. It may remarked here that although, the calculations
reported in Ref.~\cite{kiku97} are done within a quantum mechanical
theory, the point like projectile approximation is still made there.
Any how, the purpose of this figure is more to show the contribution
of the $M1$ multipolarity to the CD cross sections. As can be seen,
the contribution of this multipolarity is negligible even in the
energy bin 500-750 keV.

On the other hand, one can see from Fig. 5,
where we show the CD calculations for only $E1$ (dashed lines)
and $E2$ (dotted lines) multipolarities and
their sum (solid line) for the same reaction as in Fig. 2,
that at 250 MeV/nucleon, it is not possible to explain the data
in the region of $E_{CM}$ between 500-750 keV without the contribution
of the $M1$ multipolarity. In this figure the results obtained with
versions III (a) and IV (b) of SMEC are shown. The $E1$ + $E2$ cross
sections in version IV are somewhat larger as the $E1$ component in this
case is bigger. This sensitivity of the higher energy breakup data to the
$M1$ multipolarity makes it possible to use this to supplement the
information on the continuum structure of $^8$B which was not feasible
by similar studies at lower beam energies.  

In Fig. 6, we show the CD angular distributions ($\theta$ in this figure
corresponds to $\theta_{8_{B^*}}$) calculated with 
capture cross sections of  SMEC versions III and IV for the same
reaction as in Fig. 2. 
These results have been obtained by
integrating Eq. (\ref{cdiss}) over $E_{CM}$ between 0.1 MeV to 3.0 MeV.
The contributions of $E1$, $E2$ and $M1$ multipolarities are shown by 
by dashed, dotted and dashed-dotted curves respectively. Their sum
is shown by the solid lines. We can see that $E2$ and $M1$ contributions
start becoming important already from 1$^\circ$. Therefore, the 
requirement of the CD method that for a reliable extraction of the 
astrophysical $S$-factor the data should be dominated by the
excitation of a single multipolarity ($E1$ in present case), is more likely
to be fulfilled in the measurements at GSI energies 
if the angle $\theta_{8_{B^*}}$ is kept below 1$^\circ$.

This point is further emphasized in Fig. 7, where we show the 
energy distributions of the CD cross sections obtained by
integrating Eq. (17) for $\theta_{8_{B^*}}$ up to 1.87 $^\circ$
(angular range of the data used in this paper)
(part (a) ), and that obtained by performing the integration up
to 1$^\circ$ only (part (b) ). 

The $E1$, $E2$ and $M1$ components are
shown by dashed, dotted and dashed-dotted curves respectively,
while their sum is shown by the solid curves. The cross sections
are shown only up to $E_{CM}$ of 0.5 MeV, which is the region of
interest for the determination of the $S$-factor. It can be seen
from part (a) that if $\theta_{8_{B^*}}$ goes up to 1.87$^\circ$, the
$E2$ components are substantial (up to about 25\%) even at 
$E_{CM}$ below 0.25 MeV. 
However, if $\theta_{8_{B^*}}$ is confined to angles below
1$^\circ$, the contributions of the $E2$ component are
almost an order of magnitude 
down in comparison to those of $E1$ for $E_{CM}$ below 0.30 MeV.
Therefore, this provides a better possibility of a reliable extraction
of $S_{17}$.

\section{Summary and Outlook}

In this paper, we used the cross sections for the radiative capture 
reaction $^7$Be(p,$\gamma$)$^8$B, calculated within the shell model
embedded into the continuum approach for the structure of
$^8$B, to study the Coulomb dissociation
of $^8$B on a $^{208}$Pb target at the beam energy of 250 MeV/nucleon.
Cross sections obtained with four versions of SMEC were used.
Calculations were also performed with the capture cross sections 
obtained in a single particle model using the potential parameters
given by Esbensen and Bertsch~\cite{esbe96}.
Comparison of the calculations were made with the preliminary data 
for this reaction taken at GSI, Darmstadt recently.

The CD cross sections at these high energy were found to be 
sensitive to the nuclear structure model of $^8$B. In contrast to
the CD data taken at lower beam energies, the $M1$ multipolarity
is quite important at higher beam energies. It may therefore, 
be possible to supplement the information on the continuum structure
of $^8$B from the CD studies at higher energies. As far as the preliminary 
data are concerned, the fits obtained with SMEC approach are somewhat better
than those with the EB potential model.
We noted that if the angles of the center-of-mass of the outgoing
[$\mbox{p} - {^7}\mbox{Be}$] pair with respect to the beam direction
were taken up to 1.87$^\circ$, the $E2$ component is quite large even at 
[$\mbox{p} - {^7}\mbox{Be}$] CM energies below 0.25 MeV. To minimize the
contribution of the $E2$ multipolarity, this angle should be confined 
to values below 1$^\circ$. This conclusion appears to be by and large 
independent of the nuclear structure model of $^8$B.    

\acknowledgements
We wish to express our gratitude to S. Dro\.zd\.z,  E. Caurier 
and I. Rotter for many discussions.
We thank also F. Nowacki for stimulating collaboration in  
the course of development of the SMEC model.
This work was partly supported by
KBN Grant No. 2 P03B 097 16 and the Grant No. 76044
of the French - Polish Cooperation. One of the authors (RS) would like to
thank Abdus Salam International Centre for Theoretical Physics, Trieste,
for an associateship award.

\vfill
\newpage
\begin{table}[h]
\caption{ Parameters of the initial potentials $U(r)$ (\protect\ref{pot})
used in the calculations of self-consistent potentials $U^{(sc)}(r)$ for
two parameters $(1-\alpha)$ of the spin exchange term in the residual
interaction (\protect\ref{force}). 
All these potentials have the same parameters of radius $R_0=2.4\,$fm, surface
diffuseness $a=0.52\,$fm, and spin-orbit coupling $V_{SO}=-4\,$MeV.
 In all cases, the strength of the
residual interaction (\ref{force}) is $V_{12}^{(0)}=650\,$MeV$\cdot$fm$^{3}$
[2].}
\label{parameters}
\begin{center}
\begin{tabular}{|c|l|l|l|}
\hline
System & ${\varepsilon}_{p_{3/2}}$ [MeV] & $1-\alpha$ & $V_0$ [MeV] \\
\hline
[p $\bigotimes$ $^{7}$Be]  & $ -0.137 $ & $ 0.27 $ & $-40.045$ \\
                &            & $ 0.05 $ & $-37.660$ \\
\hline
\end{tabular}
\end{center}
\end{table}

\begin{table}[h]

\caption{ The dependence of $^8$B spectra on the relative strengths of
direct and spin exchange parts of the residual interaction
(\protect\ref{force}). 
Only ground state  of $^7$Be was taken into account in all couplings.
The proton separation energy is adjusted in order to reproduce
the energy of the lowest resonance state $1_{1}^{+}$.  
The entries in this table are labelled by
the value of the spin exchange parameter $(1 - \alpha )$ of the residual force. 
Strength of the residual interaction (\protect\ref{force}) is, 
$V_{12}^{(0)} = 650\,$MeV$\cdot$fm$^3$. 
The cut-off radius is $R_{cut}=5\,$fm except for the $p_{1/2}$ s.p.\ wave
function in $1_{1}^+$ many body states, which is in the continuum at about
$300\,$keV above the threshold and for which
larger cut-off was used $R_{cut} = 10\,$fm.
The numbers in parentheses are the
widths of $3_{1}^+$ state if this state would be
placed at the experimental energy. All units are in keV. }
\label{b8adep}
\begin{center}
\begin{tabular}{| c | r | r r | r r | r @{$\pm$} l r @{$\pm$} l |}
\hline
State & \multicolumn{1}{ c |}{SM} &
\multicolumn{2}{ c |}{$(1 - \alpha ) = 0.73$} &
\multicolumn{2}{ c |}{$(1 - \alpha ) = 0.95$} &
\multicolumn{4}{ c |}{experiment}\\
J$^{\pi}$ & energy & energy & width &  energy & width &
 \multicolumn{2}{ c }{energy} & \multicolumn{2}{ c |}{width} \\
\hline
$ 2^+ $ & $ -446 $ & $ -356 $ & --- &  $ -320 $ & --- &
 $ -137.5 $ & $ 1.0 $ & \multicolumn{2}{ c |}{ --- } \\
$ 1^+ $ & $  637 $ & $  637 $ & $ 16.5 $ &
 $ 637 $ & $ 25.9 $ & $ 637 $ & $  6 $ & $  37 $ & $ 5 $ \\
$ 3^+ $ & $ 1246 $ & $ 1294 $ & $ 13.1 $ &
 $ 1275 $ & $ 34.9 $ & $ 2183 $ & $ 30 $ & $ 350 $ & $ 40 $ \\
 & & & $ (25.2) $ & & $ (67.4) $ & \multicolumn{4}{ c |}{} \\
\hline
\end{tabular}
\end{center}
\end{table}

\newpage

\begin{figure}
\begin{center}
\mbox{\epsfig{file=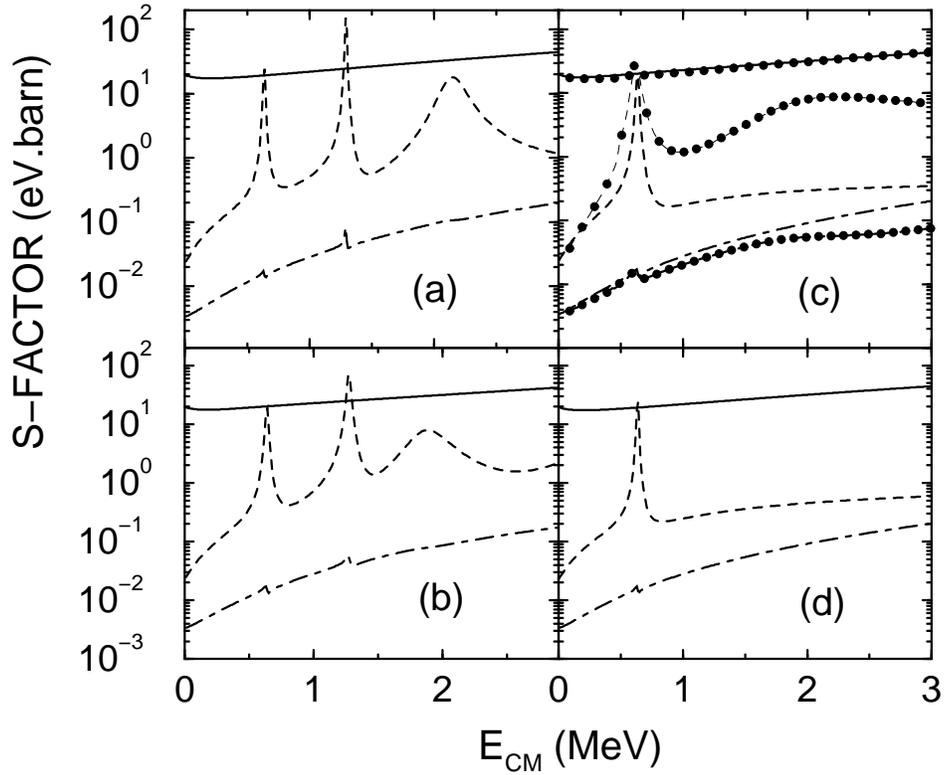,height=10.0cm}}
\end{center}
\caption[C1]{ The astrophysical S-factors for the radiative capture reaction 
$^7$Be(p,$\gamma$)$^8$B calculated with versions I, II, III and IV of
the SMEC (shown in parts (a), (b), (c) and (d) respectively). 
Solid, dashed-dotted, and dashed curves represent the 
contributions of $E1$, $E2$ and $M1$ multipolarities respectively. The
results obtained with a single particle model with potential 
parameters taken from~\protect\cite{esbe96} are also shown together 
with model III. The $E1$, $E2$ and $M1$ cross sections in this case
are shown by the solid, dashed and dotted lines with solid circles.}
\label{fig:figa}
\end{figure}

\begin{figure}
\begin{center}
\mbox{\epsfig{file=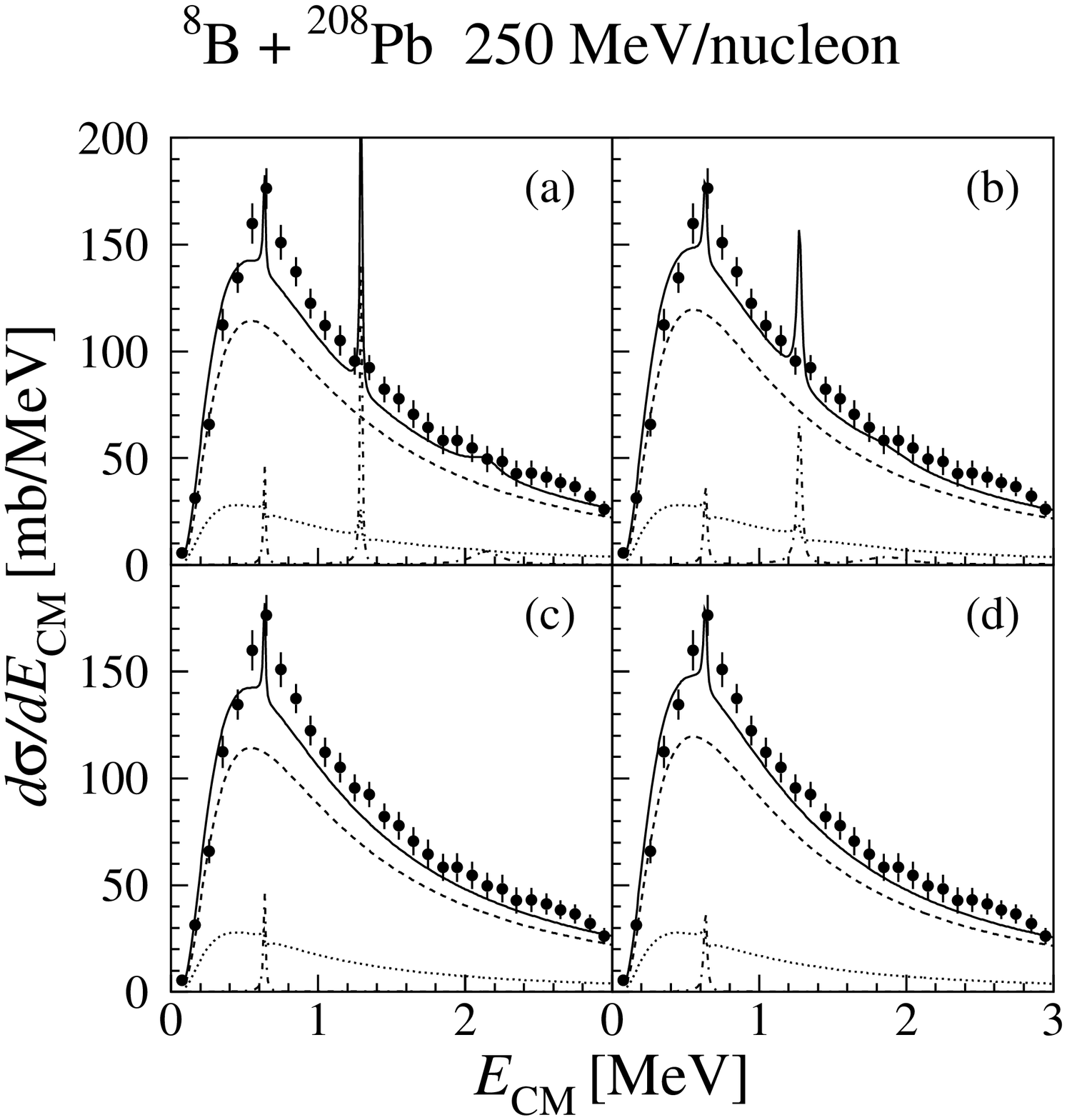,height=12.0cm}}
\end{center}
\caption[C1]{Comparison of the calculated Coulomb dissociation cross sections
($d\sigma/dE_{CM}$) with the experimental data for the breakup
of $^8$B on $^{208}$Pb target at 250 MeV/nucleon, as a function of the
$p-^7Be$ CM energy. The results calculated with four versions of
the SMEC are shown   
(models I, II, III and IV in parts {\bf (a)}, {\bf (b)}, {\bf (c)} and {\bf
(d)} respectively). Dashed, dotted and dashed-dotted curves represent the 
contributions of $E1$, $E2$ and $M1$ multipolarities respectively, while
their sum is shown by the solid line. The experimental data are taken from
the Ref.~\protect\cite{boue99}. These results have been obtained by 
integrating the double differential cross sections for angles in the  
range of 0.0$^\circ$-1.87$^\circ$. }
\label{fig:figb}
\end{figure}

\begin{figure}
\begin{center}
\mbox{\epsfig{file=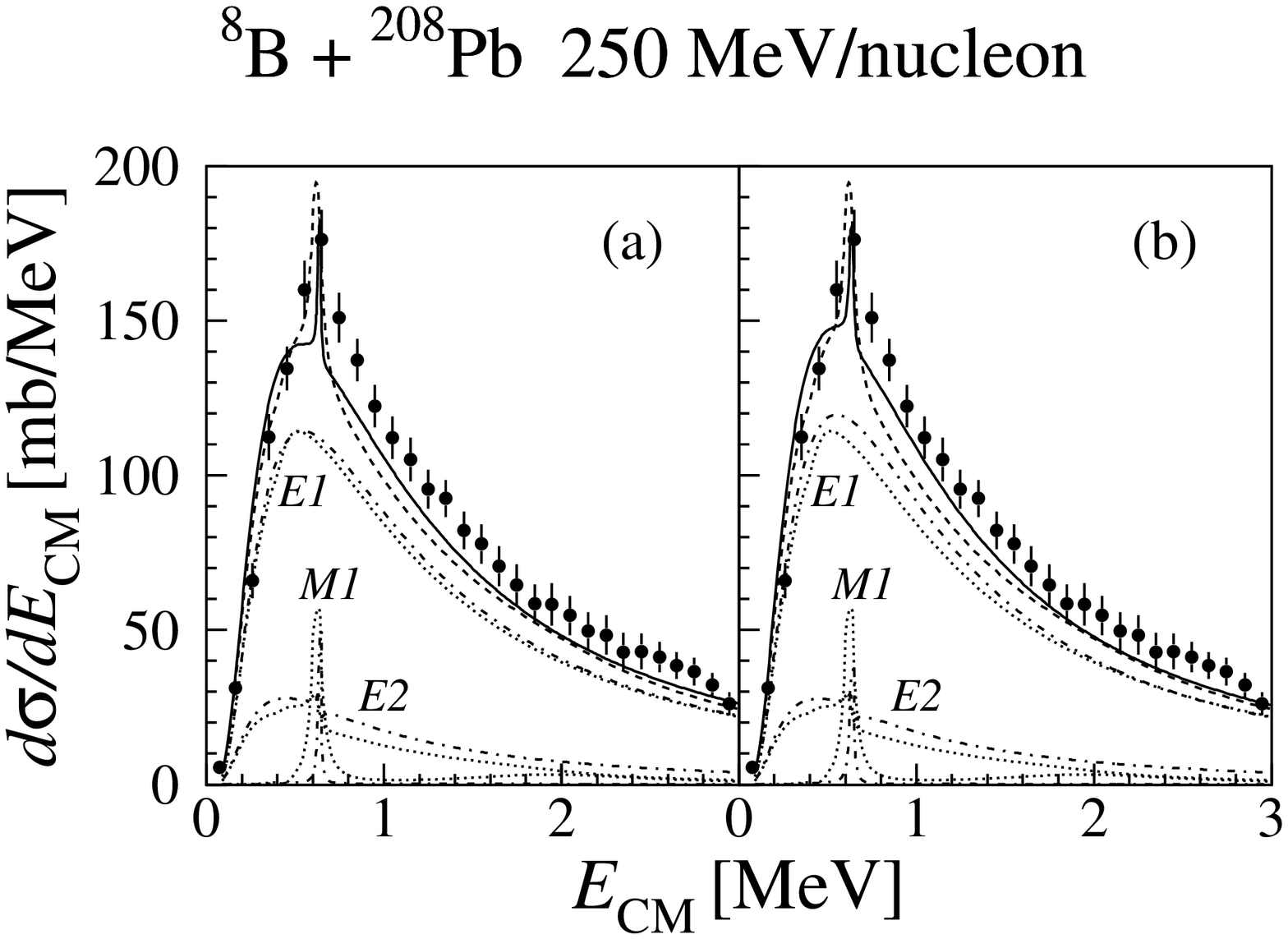,height=11.0cm}}
\end{center}
\caption{ Comparison of the CD calculations (for the same reaction and
procedure as in Fig. 1) performed with the
capture cross sections obtained with versions III (part {\bf (a)}) and IV (part
{\bf (b)}) of the SMEC with those given by the EB single particle model
\protect\cite{esbe96}. The $E1$, $E2$ and $M1$
components of the SMEC and EB models are shown by dotted and
dashed-dotted lines respectively, while the sum of these 
components are shown by solid (SMEC) and dashed (EB) lines. } 
\label{fig:figc}
\end{figure}

\begin{figure}
\begin{center}
\mbox{\epsfig{file=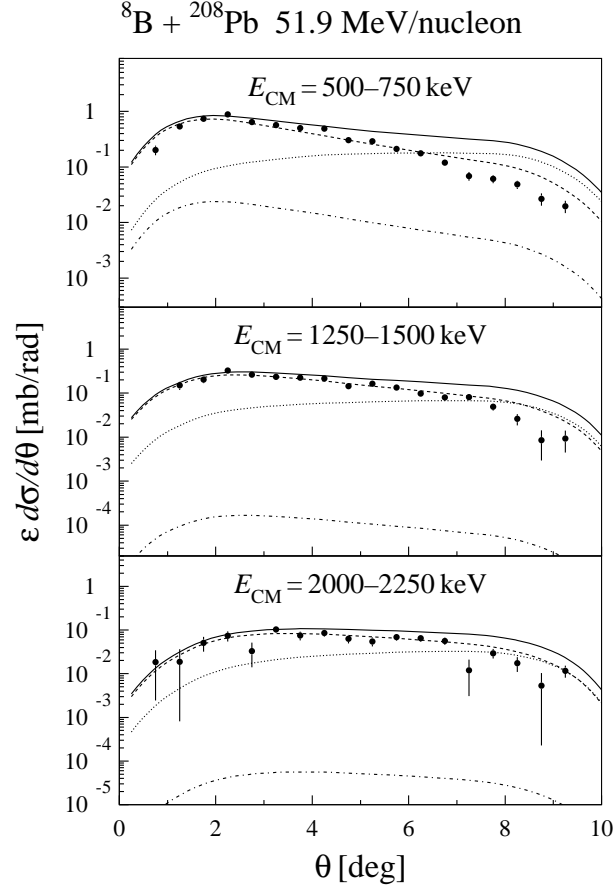,height=12.0cm}}
\end{center}
\caption{ $E1$ (dashed line), $E2$ (dotted line) and $M1$ (dashed
dotted line) components of the Coulomb dissociation cross section
$\varepsilon d\sigma/d\theta$ (calculated with version III of SMEC)
as a function of the scattering angle
$\theta$ of $^8$B$^*$ for the dissociation of $^8$B on $^{208}$Pb 
target at the beam energy of 51.9 MeV/nucleon. The solid line shows
their sum. Results for the relative energy bins of {\bf (a)}
500-750 keV, {\bf (b)} 1250-1500 keV, {\bf (c)} 2000-2250 keV are shown. 
$\varepsilon$ is the detector efficiency.
The experimental data and the detector efficiencies are taken
from~\protect\cite{kiku97}.  }
\label{fig:figd}
\end{figure}

\begin{figure}
\begin{center}
\mbox{\epsfig{file=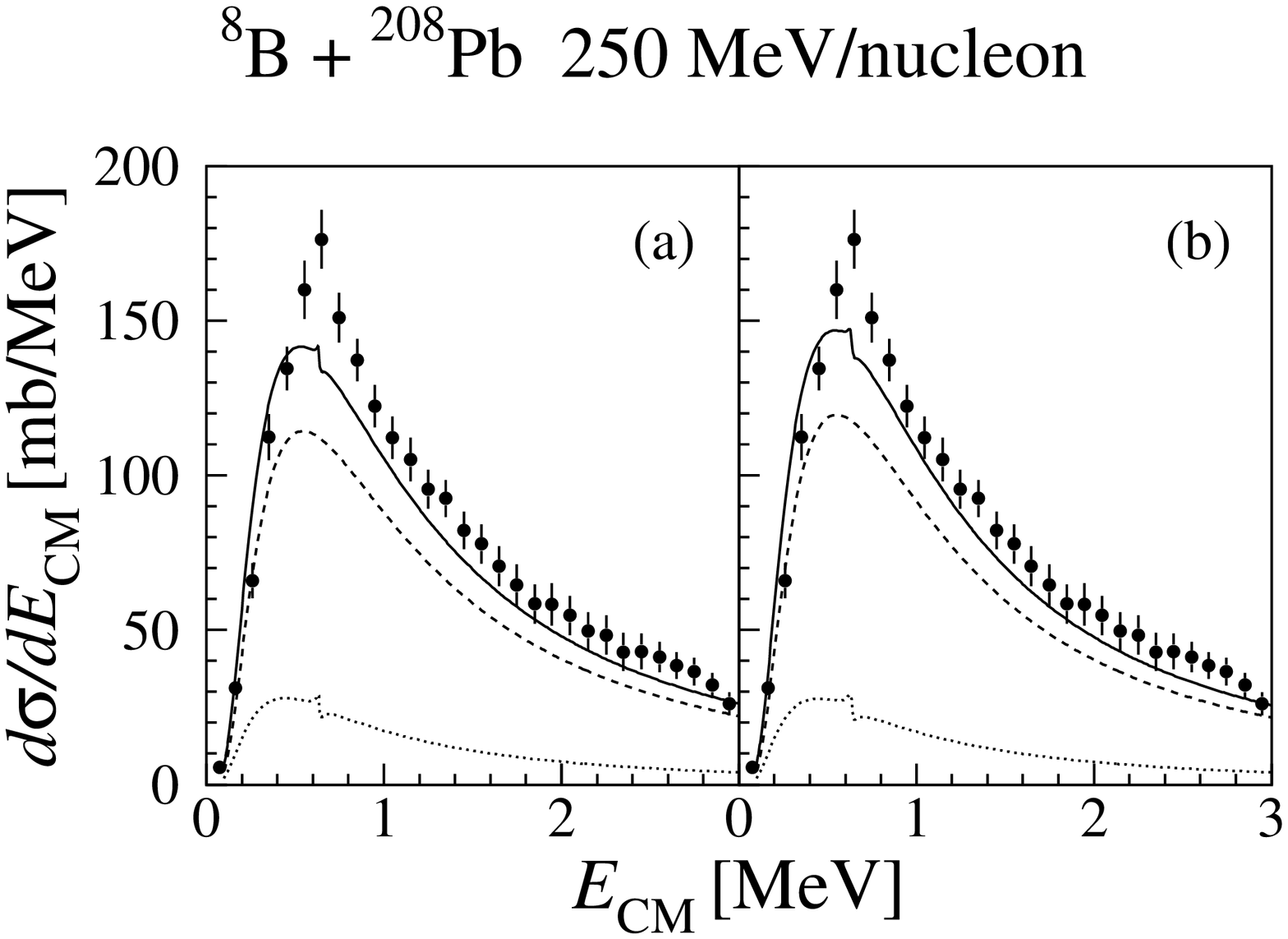,height=11.0cm}}
\end{center}
\caption[C4]{Comparison of the $E1$ + $E2$ (solid lines)
CD cross sections calculated with the capture cross sections
of versions III (part {\bf (a)}) and IV (part {\bf (b)}) of SMEC with the 
experimental data for the same reaction as in Fig. 1. The individual
$E1$ and $E2$ components are shown by dashed and dotted lines 
respectively.  }
\label{fig:fige}
\end{figure}

\begin{figure}
\begin{center}
\mbox{\epsfig{file=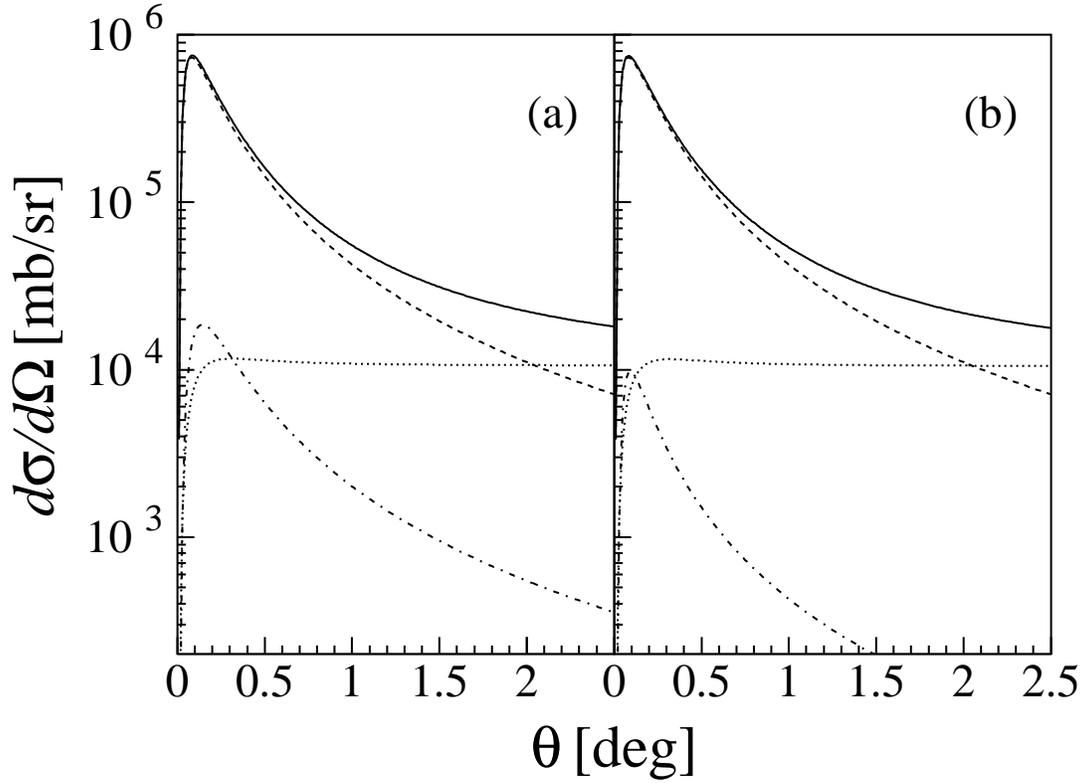,height=12.0cm}}
\end{center}
\caption[C5]{Angular distributions for the CD of $^8$B on $^{208}$Pb
at 250 MeV/nucleon calculated with versions III (part {\bf (a)}) and IV (part
{\bf (b)}) of SMEC. These results have been obtained by integrating the
double differential cross sections over the CM energies $E_{CM}$ 
between 100 keV - 3.0 MeV. }
\label{fig:figf}
\end{figure}

\begin{figure}
\begin{center}
\mbox{\epsfig{file=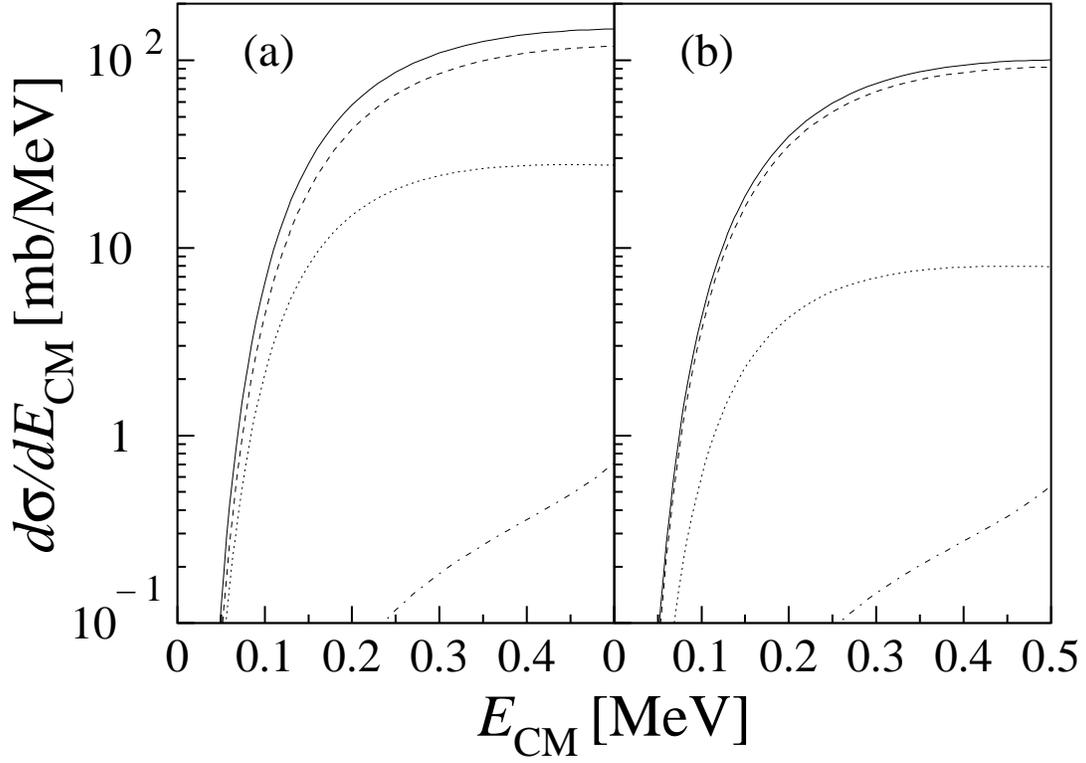,height=12.0cm}}
\end{center}
\caption[C6]{The Coulomb dissociation cross sections, $d\sigma/dE_{CM}$, 
obtianed by integrating the double differential cross sections in the
range of 0.01$^\circ$-1.87$^\circ$ (part {\bf (a)}) 
and 0.01$^\circ$-1.0$^\circ$ (part {\bf (b)}).
The $E1$, $E2$ and $M1$ components are shown by dashed, dotted and 
dashed-dotted lines respectively while their sum is depicted by the
solid lines.}
\label{fig:figg}
\end{figure}
\end{document}